# Lipkin translational-symmetry restoration in the mean-field and energy-density-functional methods

**Jacek Dobaczewski**

Institute of Theoretical Physics, Warsaw University, Hoża 69, PL-00681 Warsaw, Poland

Department of Physics, P.O. Box 35 (YFL), FI-40014 University of Jyväskylä, Finland

**Abstract.**
Based on the 1960 idea of Lipkin, the minimization of energy of a symmetry-restored mean-field state is equivalent to the minimization of a corrected energy of a symmetry-broken state with the Peierls-Yoccoz mass. It is interesting to note that the "unphysical" Peierls-Yoccoz mass, and not the true mass, appears in the Lipkin projected energy. The Peierls-Yoccoz mass can be easily calculated from the energy and overlap kernels, which allows for a systematic, albeit approximate, restoration of translational symmetry within the energy-density formalism. Analogous methods can also be implemented for all other broken symmetries.

## 1. Introduction

One of the biggest successes of the mean-field (MF) and energy-density-functional (EDF) methods applied to many-body systems consists in including correlations through the mechanism of the symmetry breaking. A single broken-symmetry MF state or a one-body broken-symmetry density can, in fact, represent a large class of correlations that are physically important. Moreover, the breaking of symmetry on the MF level corresponds to the appearance or disappearance of certain correlations that in finite systems may have all features of the phase transitions.

A link between the broken-symmetry MF states and correlated symmetry-conserving states is provided by the symmetry-restoration methods [1]. The broken-symmetry states can be viewed as auxiliary objects, which facilitate obtaining the real quantum mechanical states that have all good quantum numbers. Within the EDF methods, this leads to a very fruitful idea of the projected energy being a functional of the symmetry-breaking one-body density [2, 3]. However, variational equations then become quite difficult to solve, and the full symmetry-restored MF or EDF methods were to date applied only in several particular cases [2, 3, 4].

Numerous approximate methods to restore broken symmetries by variation before or after projection (VAP) were formulated and implemented in the past [1, 5]. The main issue here is the feasibility of the method – it must be a reasonable compromise between the precision and closeness to the exact VAP on one side, and with the numerical effort to execute it on the other side. In this respect, the Lipkin method [6] has very many advantages, which are discussed in the present study. It allows for calculating an approximate VAP energy without any necessity to perform the projection at all. In the past, it has been mostly used for the particle-number restoration, within the so-called Lipkin-Nogami [7] formulation. However, the original Lipkin method was formulated for restoring the translational symmetry, and in this study this case is studied in detail.

## 2. Results

To fix the notation, let us begin by recalling the basic standard definitions and properties pertaining to the translation symmetry. Let $|\Phi\rangle$ denote a normalized Slater determinant built of single-particle orbitals that are localized in space. Since the total momentum operator $\hat{\boldsymbol{P}} = \sum_{i=1}^{A} \hat{\boldsymbol{p}}_i$ is the generator of translation, states $|\Phi\rangle$ can be shifted by $\boldsymbol{R}$ to an arbitrary location in space as

$$|\Phi(\boldsymbol{R})\rangle = \exp(\tfrac{i}{\hbar}\boldsymbol{R}\cdot\hat{\boldsymbol{P}})|\Phi\rangle. \tag{1}$$

Then, the eigenstates of $\hat{\boldsymbol{P}}$, the so-called projected states $|\boldsymbol{P}\rangle$, can be built as linear combinations of $|\Phi(\boldsymbol{R})\rangle$, that is,

$$|\boldsymbol{P}\rangle = \frac{1}{(2\pi\hbar)^3} \int \mathrm{d}^3\boldsymbol{R}\,\exp(-\tfrac{i}{\hbar}\boldsymbol{R}\cdot\boldsymbol{P})|\Phi(\boldsymbol{R})\rangle, \tag{2}$$

and $\hat{\boldsymbol{P}}|\boldsymbol{P}\rangle = \boldsymbol{P}|\boldsymbol{P}\rangle$. The normalization condition of states $|\boldsymbol{P}\rangle$ is chosen in such a way that the original Slater determinant is a simple integral thereof,

$$|\Phi\rangle \equiv |\Phi(\boldsymbol{0})\rangle = \int \mathrm{d}^3\boldsymbol{P}|\boldsymbol{P}\rangle. \tag{3}$$

The Slater determinant $|\Phi\rangle$ is, therefore, a normalized wave packet built of non-normalizable center-of-mass plane waves $|\boldsymbol{P}\rangle$. For a system described by a translationally invariant Hamiltonian $\hat{H}$, $[\hat{H}, \hat{\boldsymbol{P}}] = 0$, one can determine the average energy $E_\Phi(\boldsymbol{P})$ of each plane wave, which is called the projected energy, as

$$E_\Phi(\boldsymbol{P}) = \langle\Phi|\hat{H}|\boldsymbol{P}\rangle/\langle\Phi|\boldsymbol{P}\rangle. \tag{4}$$

## 2.1. Kernels

From Eq. (2) one sees that the projected energy (4) is given by the Fourier transforms of the overlap and energy kernels, which are defined by

$$I(\boldsymbol{R}) = \langle\Phi|\Phi(\boldsymbol{R})\rangle, \tag{5}$$
$$H(\boldsymbol{R}) = \langle\Phi|\hat{H}|\Phi(\boldsymbol{R})\rangle, \tag{6}$$

respectively. Therefore, properties of the kernels must be discussed first. Moreover, all results below depend only on the kernels; hence these results apply automatically to the EDF approaches, where very often does not start with the Hamiltonian but the diagonal energy density is extended to the energy kernel [8].

In Figs. 1 and 2 are shown, respectively, logarithms of the overlap kernels, $\ln(I(\boldsymbol{R}))$, and reduced energy kernels, $h(\boldsymbol{R}) = H(\boldsymbol{R})/I(\boldsymbol{R})$, calculated in 9 doubly-magic spherical nuclei from $^4$He to $^{208}$Pb. Calculations were performed for the SLy4 [9] parametrization of the EDF, by using the code HFODD [10] (v2.40g) and the harmonic-oscillator (HO) basis of up to $N_0 = 14$ shells.

It can be seen that in the case of the translational symmetry, the so-called Gaussian Overlap Approximation (GOA) [1], given by

$$I(\boldsymbol{R}) = \exp(-\tfrac{1}{2}a\boldsymbol{R}^2), \tag{7}$$
$$h(\boldsymbol{R}) = h_0 - \tfrac{1}{2}h_2\boldsymbol{R}^2, \tag{8}$$

is excellent. Sudden deviations of $h(\boldsymbol{R})$ from the parabolic dependence on $\boldsymbol{R}$, which can be seen in Fig. 2 around $|\boldsymbol{R}| = 4\,\mathrm{fm}$, are due to the finiteness of the HO basis used in the calculations. Nevertheless, values of $h_2$ can be very precisely determined in the parabolic region. This was confirmed by repeating the calculations for $N_0 = 20$ HO shells, whereby the above deviations appear around $|\boldsymbol{R}| = 4.5\,\mathrm{fm}$ and the values of $h_2$ stay exactly the same.

## 2.2. Projected energies

Within the GOA, the Fourier transforms above can be analytically calculated, and the projected energy of Eq. (4) reads [1]

$$E_\Phi^{\mathrm{GOA}}(\boldsymbol{P}) = h_0 - E_{\mathrm{ZPM}} + \frac{\hbar^2\boldsymbol{P}^2}{2M_{\mathrm{PY}}}, \tag{9}$$

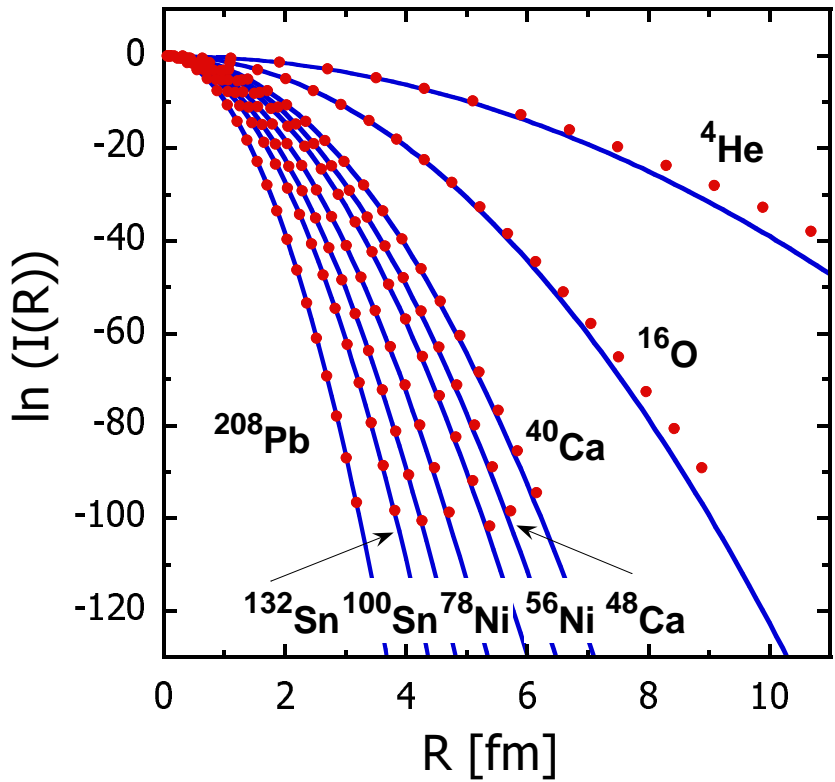

**Figure 1.** (Color online) Logarithms of the overlap kernels, $\ln(I(\bm{R}))$, calculated in 9 doubly-magic spherical nuclei (dots). Thick lines represent parabolic fits to the results, determined for $|\bm{R}| \leq 3$ fm.

where the so-called zero-point-motion correction and Peierls-Yoccoz (PY) mass are given by

$$E_{\text{ZPM}} = 3h_2/2a, \tag{10}$$
$$M_{\text{PY}} = \hbar^2 a^2/h_2. \tag{11}$$

For completeness, in Figs. 3 and 4, I show the mass dependence of the GOA parameters $a$, $h_2$, $E_{\text{ZPM}}$, and $M_{\text{PY}}$, along with the fits of the power-law dependencies. One clearly sees that the PY mass is not equal to the total mass $mA$ of the nucleus. However, as discussed below, it is the PY mass, and not the physical mass $mA$, which is important for the translational-symmetry restoration.

The average energy of the system at rest, $E_\Phi(\bm{0})$, depends on the Slater determinant $|\Phi\rangle$, and in what follows we are interested in minimizing this energy with respect to $|\Phi\rangle$, that is, in performing the VAP or variation after symmetry restoration. To this end, in this study I follow the seminal idea by Harry Lipkin [6], who realized that the VAP calculations can be very easily performed by flattening the function $E_\Phi(\bm{P})$.

Indeed, from Eqs. (3) and (4) one obtains the so-called sum-rule property,

$$\int d^3\bm{P}\, E_\Phi(\bm{P})\langle\Phi|\bm{P}\rangle = \langle\Phi|\hat{H}|\Phi\rangle, \tag{12}$$

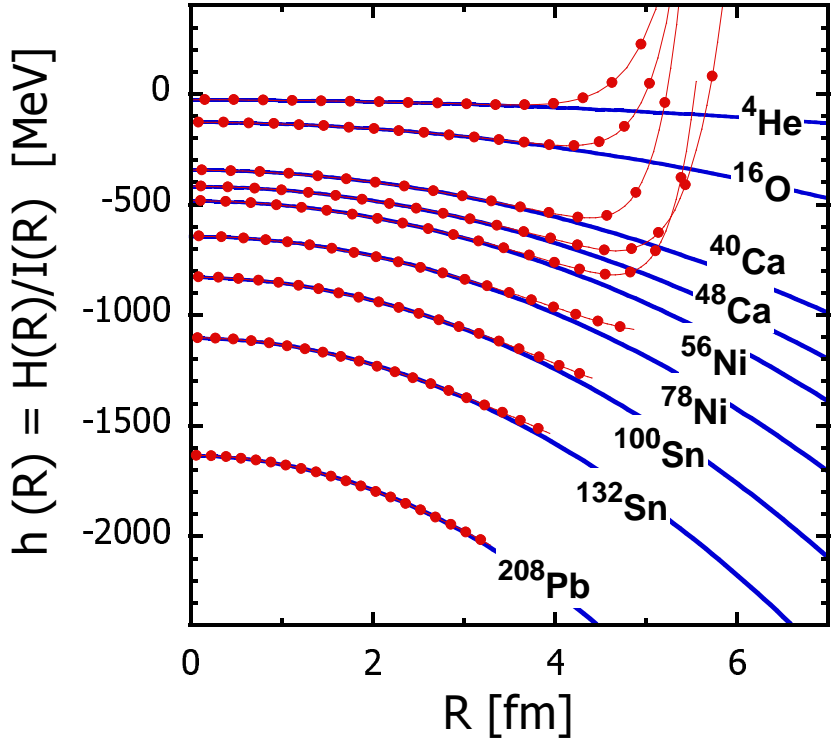

**Figure 2.** (Color online) Same as in Fig. 1 but for the reduced energy kernels, $h(\boldsymbol{R}) = H(\boldsymbol{R})/I(\boldsymbol{R})$. To guide the eye, thin lines connect calculated values (dots).

which tells us that the MF energy $\langle\Phi|\hat{H}|\Phi\rangle$ is equal to the average of the projected energies $E_\Phi(\boldsymbol{P})$ weighted by amplitudes $\langle\Phi|\boldsymbol{P}\rangle$. Therefore, minimization of $\langle\Phi|\hat{H}|\Phi\rangle$, that is, the standard MF method, corresponds to an entangled minimization of both the projected energies $E_\Phi(\boldsymbol{P})$ and amplitudes $\langle\Phi|\boldsymbol{P}\rangle$, whereas the VAP method pertains to the minimization of the energy $E_\Phi(\boldsymbol{0})$ only, and, of course, disregards the amplitudes $\langle\Phi|\boldsymbol{P}\rangle$ completely.

Note that the GOA amplitudes $\langle\Phi|\boldsymbol{P}\rangle$ are all positive,

$$\langle\Phi|\boldsymbol{P}\rangle = \left(\frac{2\pi a}{\hbar^2}\right)^{-3/2} \exp\left(-\frac{\hbar^2}{2a}\boldsymbol{P}^2\right), \quad (13)$$

and hence the MF minimization is bound to underestimate the momentum spread of the Slater determinant. It is then obvious that unless the center-of-mass and internal degrees of freedom are exactly separated, like is the case for the closed-shell HO [11, 12, 13] or for the coupled-cluster states [14], the MF and VAP Slater determinants can be different.

### 2.3. The Lipkin method

The Lipkin idea of flattening the function $E_\Phi(\boldsymbol{P})$ is implemented in the following way. Guided, by the GOA result in Eq. (9), one defines the Lipkin operator

$$\hat{K} = k\hat{\boldsymbol{P}}^2, \quad (14)$$

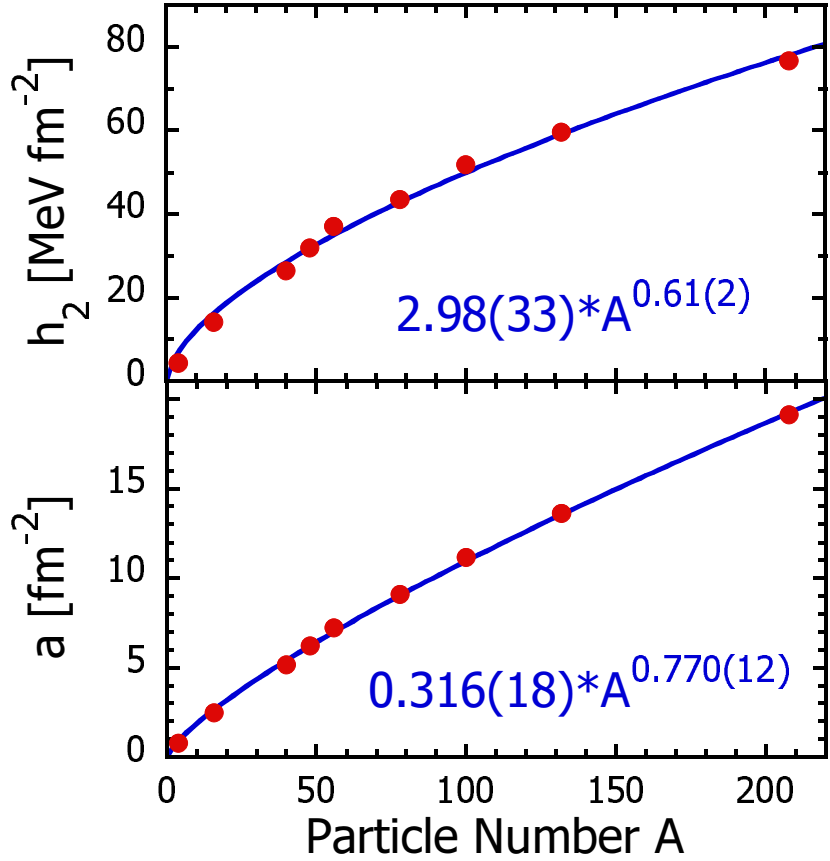

**Figure 3.** (Color online) Dots show parameters $a$ (bottom) and $h_2$ (top) that are calculated by the parabolic fits to the logarithms of the overlap kernels (Fig. 1) and reduced energy kernels (Fig. 2). Lines give estimates of the power-law dependence.

for which the projected average values $K(\boldsymbol{P})$ are defined as:

$$K(\boldsymbol{P}) = \langle\Phi|\hat{K}|\boldsymbol{P}\rangle/\langle\Phi|\boldsymbol{P}\rangle, \tag{15}$$

which in turn fulfill the sum rule:

$$\int \mathrm{d}^3\boldsymbol{P}\, K(\boldsymbol{P})\langle\Phi|\boldsymbol{P}\rangle = \langle\Phi|\hat{K}|\Phi\rangle. \tag{16}$$

Then, by subtracting Eqs. (12) and (16), one obtains

$$\int \mathrm{d}^3\boldsymbol{P}\, (E_\Phi(\boldsymbol{P}) - K(\boldsymbol{P}))\,\langle\Phi|\boldsymbol{P}\rangle = \langle\Phi|\hat{H} - \hat{K}|\Phi\rangle. \tag{17}$$

It is clear that the flattest difference $E_\Phi(\boldsymbol{P})-K(\boldsymbol{P})$ is obtained by adjusting the constant $k$ so as to best fulfill the equation

$$K(\boldsymbol{P}) = E_\Phi(\boldsymbol{P}) - E_\Phi(\boldsymbol{0}). \tag{18}$$

Note that since $K(\boldsymbol{0}) = 0$, the exact projected energy of the system at rest, $E_\Phi(\boldsymbol{0})$, must, by definition, appear on the right-hand side of Eq. (18).

In case that $E_\Phi(\boldsymbol{P})$ grows exactly parabolically, one obtains that

$$E_\Phi(\boldsymbol{0}) = \langle\Phi|\hat{H} - \hat{K}|\Phi\rangle, \tag{19}$$

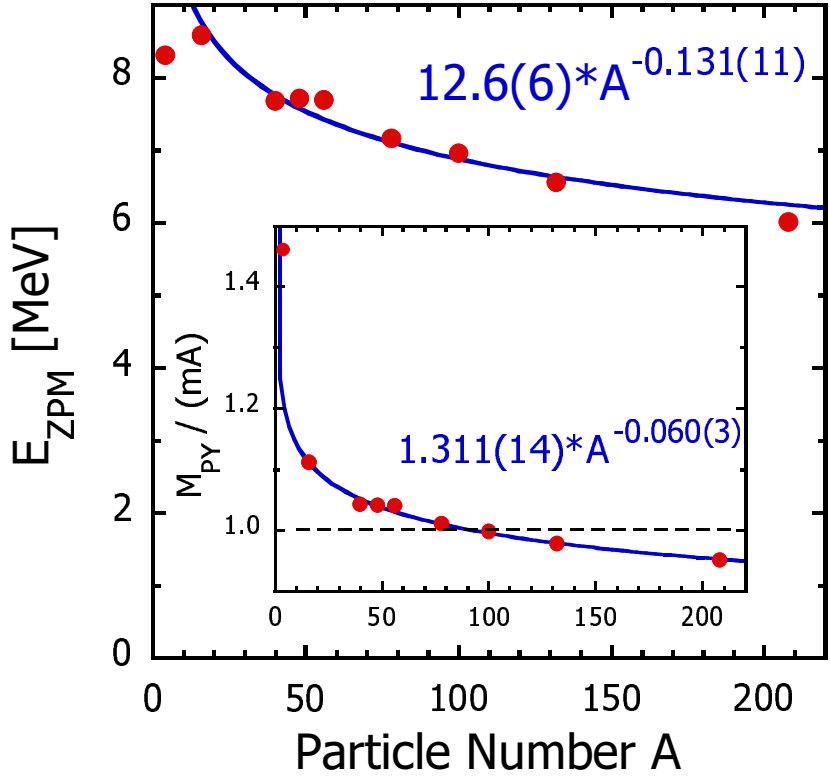

**Figure 4.** (Color online) The GOA zero-point-motion corrections $E_{\text{ZPM}}$ and PY masses $M_{\text{PY}}$ (inset) calculated for the values of parameters $a$ and $h_2$ that are shown in Fig. 3.

that is, the projected energy of the system at rest $E_\Phi(\mathbf{0})$ can be calculated without performing any projection at all.

The essence of the Lipkin method is in finding a suitable value of the correcting parameter $k$, which for each Slater determinant $|\Phi\rangle$ must describe the parabolic growth of the function $E_\Phi(\mathbf{P})$. We note here that this growth has nothing to do with the physical translational motion of the system boosted to momentum $\mathbf{P}$, in which case the energy must grow as $E(\mathbf{P}^2) = \hbar^2 \mathbf{P}^2/2mA$. The function $E_\Phi(\mathbf{P})$ simply characterizes the distribution of projected energies within the Slater determinant, that is, the degree of the symmetry breaking in the Slater determinant at rest. Therefore, the correcting parameter $k$ has no obvious relation with the true translational mass of the system. Moreover, the correcting parameter must depend on all kinds of approximations or space truncations, which are made when obtaining the Slater determinant $|\Phi\rangle$, that is, the Lipkin method corrects for these approximations too.

The right-hand side of Eq. (19) can be minimized by the standard self-consistent method, whereby the optimum state $|\Phi\rangle$ can be found. At each stage of the iterative procedure one has to determine $k$, that is, $k$ must parametrically depend on $|\Phi\rangle$. Note that at any given iteration of the self-consistent method, this parametric dependence *must not be varied*. Finally, after the iteration converges, $E_\Phi(\mathbf{0})$ is given by the obtained minimum value of the Lipkin projected energy (19). Obviously, the quality of the result

crucially depends on the quality of the calculation of $k$, which now will be discussed.

### 2.4. The Lipkin operator

A practical method to calculate the correcting constant $k$ from Eq. (18) must not involve, of course, the exact evaluation of the function $E_\Phi(\boldsymbol{P})$. A method to probe the function $E_\Phi(\boldsymbol{P})$ without evaluating it explicitly can be formulated as follows. One first remarks that the projected energy (4) can also be calculated as:

$$E_\Phi(\boldsymbol{P}) = \langle\Phi|\hat{H}\hat{g}_n(\hat{\boldsymbol{P}})|\boldsymbol{P}\rangle/\langle\Phi|\hat{g}_n(\hat{\boldsymbol{P}})|\boldsymbol{P}\rangle, \tag{20}$$

which gives the sum rules:

$$\int \mathrm{d}^3\boldsymbol{P}\, E_\Phi(\boldsymbol{P}) g_n(\boldsymbol{P}) \langle\Phi|\boldsymbol{P}\rangle = \langle\Phi|\hat{H}\hat{g}_n(\hat{\boldsymbol{P}})|\Phi\rangle, \tag{21}$$

where $\hat{g}_n(\hat{\boldsymbol{P}})$ are arbitrary functions of $\hat{\boldsymbol{P}}$. By using different functions $\hat{g}_n$, one can probe the unknown function $E_\Phi(\boldsymbol{P})$.

An obvious choice of $\hat{g}_n = \hat{\boldsymbol{P}}^{2n}$, which, in fact, has been used within the Lipkin-Nogami method [7] to restore the particle number, requires dealing with impractical many-body operators. A much better option is provided by the shift operators [cf. Eq. (1) and (2)],

$$\hat{g}_n = \exp(\tfrac{i}{\hbar}\boldsymbol{R}_n \cdot \hat{\boldsymbol{P}}), \tag{22}$$

defined for a suitably selected values of shifts $\boldsymbol{R}_n$. Then, the average values on the right-hand side of Eq. (21) become equal to the energy kernels $H(\boldsymbol{R}_n)$, which are very easy to calculate.

In practice, the method works as follows. Suppose one wants to evaluate the Taylor-expansion coefficients of $E_\Phi(\boldsymbol{P})$ up to a given order of $2M$,

$$E_\Phi(\boldsymbol{P}) = \sum_{m=0}^{M} E_\Phi^{(2m)} \boldsymbol{P}^{2m}. \tag{23}$$

After inserting this into Eq. (21), one obtains the set of linear equations,

$$\sum_{m=0}^{M} A_{nm} E_\Phi^{(2m)} = H(\boldsymbol{R}_n). \tag{24}$$

where the matrix $A_{nm}$ is defined by the kernels of the momentum operators:

$$A_{nm} = P_{2m}(\boldsymbol{R}_n) \equiv \langle\Phi|\hat{\boldsymbol{P}}^{2m}|\Phi(\boldsymbol{R}_n)\rangle. \tag{25}$$

In the simplest case of the quadratic Lipkin operator, Eq. (14), one only needs the expansion up the second order, that is, for $M = 1$. Then, by using two points $\boldsymbol{R}_0 \equiv \boldsymbol{0}$ and $\boldsymbol{R}_1 \equiv \boldsymbol{R}$ one obtains

$$A = \begin{pmatrix} 1 & , & \langle\Phi|\hat{\boldsymbol{P}}^2|\Phi\rangle \\ \langle\Phi|\Phi(\boldsymbol{R})\rangle & , & \langle\Phi|\hat{\boldsymbol{P}}^2|\Phi(\boldsymbol{R})\rangle \end{pmatrix}. \tag{26}$$

This matrix can be easily inverted and then one obtains the first two Taylor expansion coefficients $E_\Phi^{(0)}$ and $E_\Phi^{(2)}$, which are required in Eq. (18), that is,

$$E_\Phi(\mathbf{0}) = \frac{h(\mathbf{0})p_2(\mathbf{R}) - h(\mathbf{R})p_2(\mathbf{0})}{p_2(\mathbf{R}) - p_2(\mathbf{0})}, \qquad (27)$$

$$k = \frac{h(\mathbf{R}) - h(\mathbf{0})}{p_2(\mathbf{R}) - p_2(\mathbf{0})}, \qquad (28)$$

where the reduced kernel of the momentum operator is defined as usually, by $p_2(\mathbf{R}) = P_2(\mathbf{R})/I(\mathbf{R})$. Expressions (27) and (28) can be very easily evaluated, especially in view of the fact that the momentum kernel can be calculated as a Laplacian of the overlap kernel:

$$P_2(\mathbf{R}) = -\hbar^2 \Delta_\mathbf{R} I(\mathbf{R}). \qquad (29)$$

Of course, within the GOA, the results are exactly the same as those given by the zero-point-motion correction and PY mass, Eqs. (10) and (11). However, expressions (27) and (28) do not rely on the GOA. They only depend on assuming the quadratic form of the Lipkin operator $\hat{K}$, Eq. (14). Moreover, variations of explicit expressions for the projected energy $E_\Phi(\mathbf{0})$, like the ones given by Eqs. (9) or (27), are difficult, while that of the Lipkin projected energy (19) can be carried out by the standard self-consistent procedure.

When the quadratic approximation is not sufficient, one can immediately notice this fact by a dependence of $E_\Phi(\mathbf{0})$ and $k$ on the value of the shift $\mathbf{R}$. In this case, one can always switch to higher-order Lipkin operators:

$$\hat{K} = \sum_{m=1}^{M} k_{2m} \hat{\mathbf{P}}^{2m}, \qquad (30)$$

which would require using higher-order Taylor expansions (23), and $M$ different shifts $\mathbf{R}_n$, $n = 1, \ldots, M$, instead of one. The only requirement for choosing the shifts $\mathbf{R}_n$ is a non-singularity of the matrix $A$. A dependence of the results on this choice will always give a signal that a given order is insufficient. Note that kernels of higher powers of the momentum operator can also be calculated in terms of higher derivatives of the overlap kernel, in analogy with Eq. (29). However, in view of the fact that for the translational symmetry the GOA works so nicely, in this study there does not seem to be any immediate necessity to go to higher orders, and the simple quadratic correction will suffice.

### 2.5. The direct part of $\langle \Phi | \hat{\mathbf{P}}^2 | \Phi \rangle$

Standard calculations for the SLy4 Skyrme functional [9] are performed for the Lipkin operator of $\hat{K} = \hat{T}/A$, which is a fixed factor of $A$ smaller than the one-body kinetic-energy operator $\hat{T}$. This procedure simply renormalizes the single particle masses as $m' = m(1-1/A)$ [15]. Another standard was adopted for the SLy6 and SLy7 functionals, whereby the Lipkin operator of Eq. (14) with the exact mass, that is $k = \hbar^2/2mA$, was used. In Fig. 5, are compared the Lipkin projected energies calculated with the exact and PY, Eq. (28), masses and Fig. 6 shows values of these masses. In all cases, the PY masses were calculated by using in Eq. (14) the shift of $|\boldsymbol{R}| = 2\,\text{fm}$.

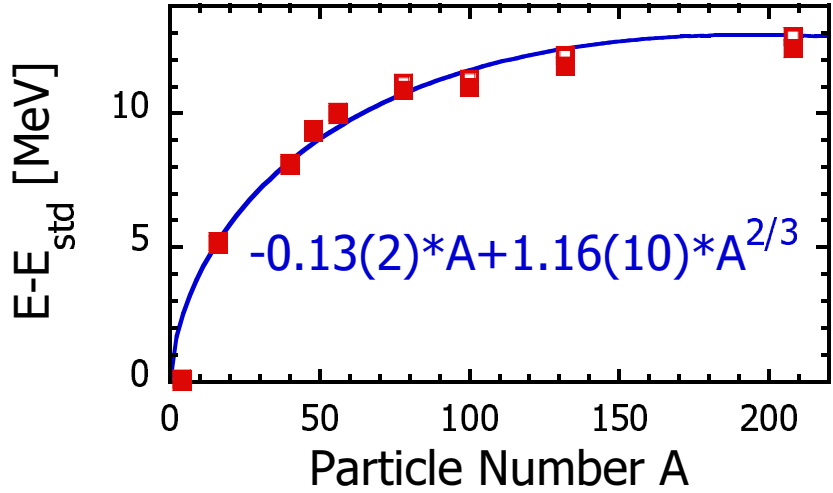

**Figure 5.** (Color online) The Lipkin projected energies (19) calculated in 9 doubly-magic spherical nuclei, and plotted relative to the standard energies calculated for the SLy4 Skyrme functional [9]. Open and full squares show results obtained for the exact $[k = \hbar^2/2mA]$ and calculated [Eq. (28)] correcting factors, respectively. Solid line shows the fit of the volume and surface terms.

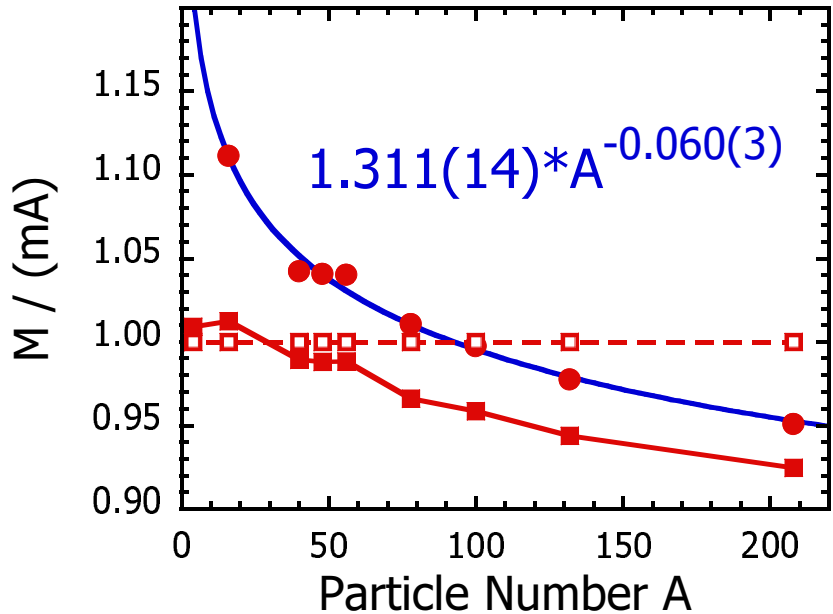

**Figure 6.** (Color online) Exact masses ($M = mA$, open squares) and the PY masses calculated after the Lipkin minimization from Eq. (28) ($M = \hbar^2/2k$, full squares), compared with the PY masses of Fig. 4 (full circles).

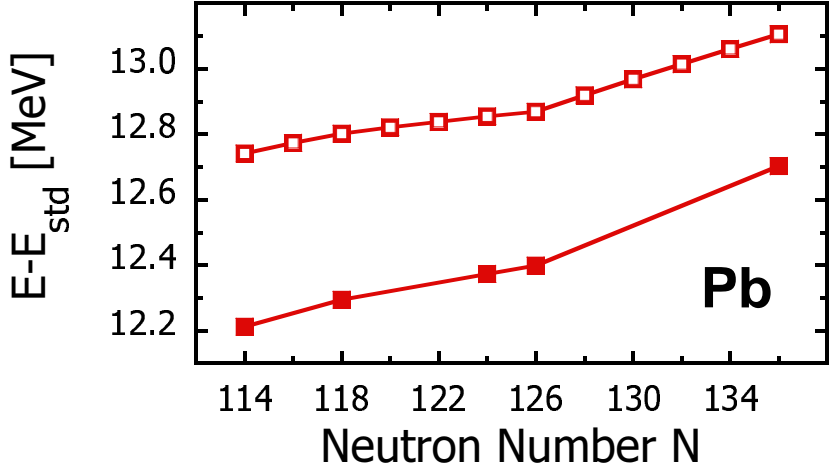

**Figure 7.** (Color online) Same as in Fig. 5 but for the chain of lead isotopes around $^{208}$Pb.

First one sees that the Lipkin projected energies obtained with the two-body Lipkin operators, Eq. (14), differ from those using the standard one-body operator $\hat{T}/A$ by up to 13 MeV. The mass dependence of this difference can be very well described by the sum of the volume and surface terms. Therefore, its major part can easily be absorbed in the parameters of the Skyrme functional. This confirms conclusions of Ref. [16]. Second, one sees in Fig. 6 that the values of the PY masses differ form the exact ones only by a few percent. Therefore, the Lipkin projected energies based on these two prescriptions for the normalizing constant $k$ differ very little.

Results shown in Fig. 7 aim at checking whether the differences between the exact-mass and PY-mass correcting factors can influence shell effects in masses of lead isotopes. One sees, that these differences induce an almost constant shift of the energy, of the order of 0.5 MeV. Moreover, the shell effect at $^{208}$Pb, induced by replacing the Lipkin operator $k\hat{\boldsymbol{P}}^2$ by $\hat{T}/A$, is very small.

Therefore, based on these results, one might be tempted to consider $\hat{T}/A$ as a viable alternative to $k\hat{\boldsymbol{P}}^2$. However, in view of the Lipkin symmetry-restoration method, this is not the case. This is shown in Fig. 8, where are compared the GOA properties of the reduced kernels of $\hat{\boldsymbol{P}}^2$ and $\hat{T}$,

$$p_2(\boldsymbol{R}) = \frac{\langle\Phi|\hat{\boldsymbol{P}}^2|\Phi(\boldsymbol{R})\rangle}{\langle\Phi|\Phi(\boldsymbol{R})\rangle} = p_{20} - \tfrac{1}{2}p_{22}\boldsymbol{R}^2. \qquad (31)$$

$$\pi(\boldsymbol{R}) = \frac{\langle\Phi|\frac{2m}{\hbar^2}\hat{T}|\Phi(\boldsymbol{R})\rangle}{\langle\Phi|\Phi(\boldsymbol{R})\rangle} = \pi_0 - \tfrac{1}{2}\pi_2\boldsymbol{R}^2, \qquad (32)$$

where the factor in front of the kinetic-energy operator was chosen so that $\pi$ is simply the kernel of the direct part of the operator $\hat{\boldsymbol{P}}^2$ [16]. Again, the quality of the GOA is here excellent, so the parameters $p_{22}$ and $\pi_2$ characterize the kernels very well, while the parameters $p_{20} = \langle\Phi|\hat{\boldsymbol{P}}^2|\Phi\rangle$ and $\pi_0 = \langle\Phi|\frac{2m}{\hbar^2}\hat{T}|\Phi\rangle$ give the standard average values.

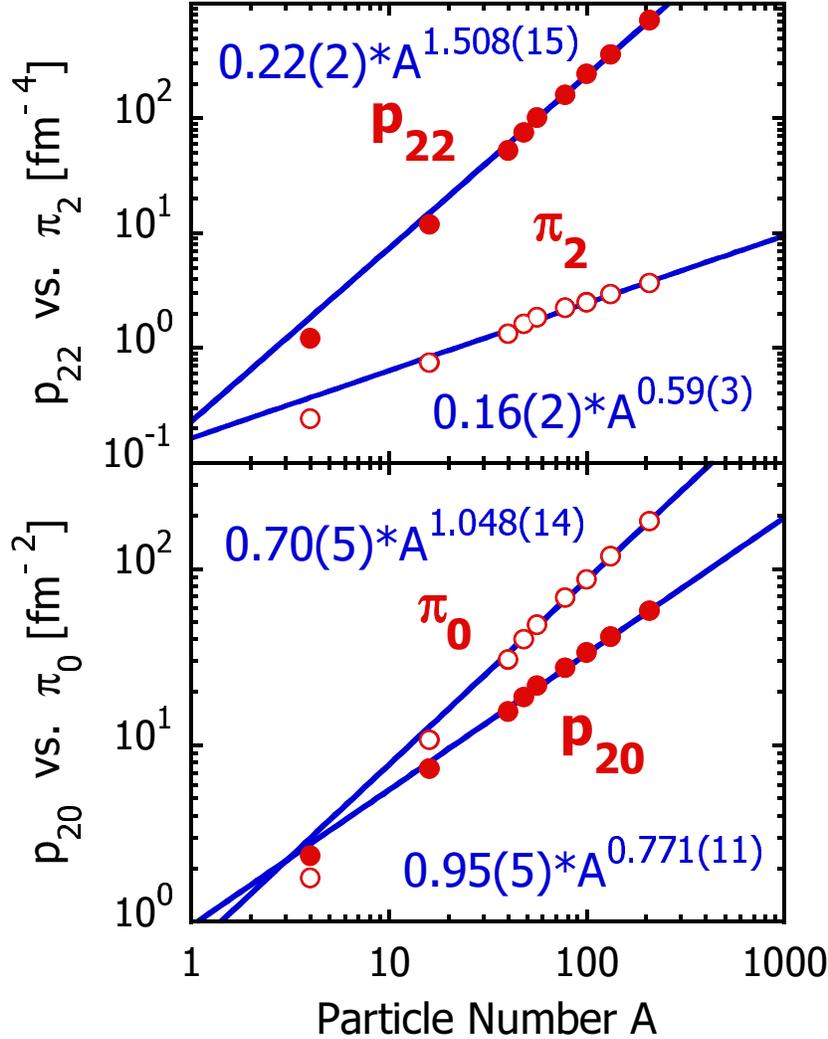

**Figure 8.** (Color online) Mass-number dependence of the 0-th (lower panel) and 2-nd (upper panel) GOA moments of the two-body momentum operator squared and its direct part, see Eqs. (31) and (32).

In the bottom panel of Fig. 8, one sees that the direct parts ($\pi_0$) increase almost linearly with the particle number $A$, while the true average values ($p_{20}$) increase more like $A^{3/4}$. In a heavy nucleus like $^{208}$Pb, these trends result in $p_{20}$ being overestimated by $\pi_0$ by about a factor of 3, which confirms the results obtained in Ref. [17]. Such mass dependencies explain why this error can be fairly well absorbed within the volume ($\sim A$) and surface ($\sim A^{2/3}$) energies of the Skyrme functional.

However, in the top panel of Fig. 8, one sees that the second moments of these kernels, $p_{22}$ and $\pi_2$, behave quite differently. The second moment of the true kernel ($p_{22}$) increases quite fast, like $A^{3/2}$, while the direct part ($\pi_2$) only as $A^{1/2}$. In $^{208}$Pb, these trends result in $p_{22}$ being underestimated by $\pi_2$ by about a factor of 200. It is then obvious that one cannot replace the true average values $\langle\Phi|\hat{\boldsymbol{P}}^2|\Phi\rangle$ by the direct parts, no matter which value of the multiplicative factor is used.

One can note here in passing that, within the GOA, Eq. (29) gives the relation $p_{22} = \frac{2}{9}p_{20}^2$, which is very well fulfilled by the numerical results shown in Fig. 8. Than, again within GOA, the average values of $\hat{T}$ in the projected states can be calculated in analogy to Eq. (9), and are proportional to

$$\pi^{\text{GOA}}(\boldsymbol{P}) = \pi_0 - \frac{3\pi_2}{2a} + \frac{\pi_2}{2a^2}\boldsymbol{P}^2, \qquad (33)$$

The fact that $\pi_2$ is a factor of 200 too small in $^{208}$Pb means that the function $\pi(\boldsymbol{P})$ is not able to flatten the $\boldsymbol{P}^2$ dependence of $E_\Phi(\boldsymbol{P})$, and thus the method based on replacing $k\hat{\boldsymbol{P}}^2$ by $\hat{T}/A$ does not lead to a proper VAP estimate of the projected energy.

## 3. Conclusions and perspectives

In the present study, I have analyzed the translational-symmetry restoration by the approximate variation after projection based on the Lipkin method. For translational symmetry, the Gaussian Overlap Approximation gives an excellent representation of numerical results obtained in doubly-magic spherical nuclei by using the Skyrme energy density functionals. The Lipkin method is based on subtracting from the Skyrme energy the center-of-mass kinetic energy with the Peierls-Yoccoz mass, and not with the true mass. However, calculations show that the Peierls-Yoccoz mass is only a few percent different than the true mass of the nucleus.

I have also studied properties of the direct part of the center-of-mass kinetic energy and I showed that this direct part does not fulfill Lipkin conditions and thus is not a proper correction within the variation after projection method to restore the translational symmetry. In fact, it seems that there is no good argument in support of neglecting the exchange part of the center-of-mass kinetic energy at all, apart from the fact that it is a cheap method. Nevertheless, shell effects induced by the exchange part appear to be weak, and thus the exchange part may only weakly influence the overall agreement of the calculated masses with experimental data.

Due to the fact that the three Cartesian components of the total momentum operator commute, the Lipkin method to restore the translational symmetry can be easily generalized to deformed nuclei. In this case, one obtains two or three different Peierls-Yoccoz translational masses in axially or triaxially deformed nuclei, respectively. This natural result reflects the fact that in deformed nuclei the momentum fluctuations along the principal axes of the mass distribution are different.

The methods presented in this study may constitute an interesting alternative to the Lipkin-Nogami expressions [7, 18] for calculating the correcting factor $\lambda_2$, required for the approximate particle-number-symmetry restoration. Indeed, the Lipkin-Nogami expressions are difficult to implement in realistic calculations, and approximate work-around procedures have been used in practical approaches [19, 4]. Expressions analogous to Eqs. (28) and (29), which are based on the transition energy and overlap kernels with respect to a simple shift in the gauge angle, could open here a quite useful new possibility.

The largest impact of the Lipkin method can be expected for the rotational-symmetry restoration. Here, the Lipkin operator corresponds to the total angular momentum squared, which, being included before variation in the functional, may induce deformation in all nuclei, including the magic ones. This may create substantial rotational corrections in these nuclei, in analogy to those that are obtained by minimizing the angular-momentum projected energies after variation over deformation, see e.g. Ref. [20]. However, the Lipkin method may magnify these corrections even more, because it treats them before variation. Since in the case of the rotational symmetry, the Lipkin corrections can be expected to be strongly shell dependent, they may have a strong impact on the agreement of the calculated masses with experimental data.

The same methodology can also be applied to restore the isospin symmetry in nuclei. Here, the Lipkin operator corresponds to the total isospin operator squared, cf. Refs. [21, 22], and the Peierls-Yoccoz mass must be calculated by considering kernels of the nuclear energy only. In this way, the dependence of this nuclear energy on the total isospin is flattened, which this leaves to the Coulomb energy the possibility of inducing the correct isospin-mixing effects.

It is also worth noting that the Lipkin method with the Peierls-Yoccoz collective mass is based on properties of the energy kernels for relatively small shifts of the collective coordinates. Therefore, singularities of these kernels, see Ref. [23] and references cited therein, which plague the exact symmetry-restoration methods within the energy-density-functional approaches, are not causing problems. This observation has to be substantiated by analyzing the singularity-free corrected kernels [24, 25, 26] and checking that the proposed corrections are appropriately small for small shifts of the collective coordinates.

Altogether, the Lipkin method reviewed in this study can be systematically applied to restore, within the mean-field or energy-density-functional theories, all broken symmetries. This method constitutes a practical alternative with respect to the exact projection techniques, which are very costly and thus cannot be applied to several broken symmetries simultaneously. The work on implementing the Lipkin method to symmetries other than the translational one studied in this work is now in progress.

Finally, the Lipkin method can also be used for approximate calculation of average values of observables in projected states. However, this method cannot replace the real projection in cases when the good quantum numbers are necessary to calculate matrix elements and properly account for selection rules. Nevertheless, by projecting good quantum numbers from the state that minimizes the Lipkin projected energy one probably obtains the best viable alternative to the exact variation after projection.

This work was supported in part by the Polish Ministry of Science, by the Academy of Finland and University of Jyväskylä within the FIDIPRO programme, and by the UNEDF SciDAC Collaboration under the U.S. Department of Energy grant No. DE-FC02-07ER41457.